\documentclass[11pt]{article}
\usepackage{graphicx}
\usepackage{epsfig}
\usepackage{times}
\begin{document}
\pagestyle{empty}
\begin{center}
{\bf \Large MOLECULAR DYNAMICS STUDY OF THE\\
            GLASS TRANSITION IN CONFINED WATER}
\vskip 1cm
P.~Gallo and  
M.~Rovere
\vskip 0.5cm
Dipartimento di Fisica, Universit{\`a} di Roma
Tre, and\\
Istituto Nazionale per la Fisica della Materia,  Unit{\`a} di Ricerca
Roma Tre, \\ Via della Vasca Navale 84, I-00146 Roma, Italy \\

\bigskip
\noindent
\end{center}

\begin{flushright}
\parbox[t]{16.5truecm}
{{\footnotesize{\bf Abstract.}
A molecular dynamics simulation of SPC/E water confined in a 
Silica pore is presented. The pore has been constructed to 
reproduce the average properties of a pore of Vycor glass. 
Due to the confinement and to the
presence of a strong hydrophilic surface,
the dynamic behaviour of the liquid appears to be 
strongly dependent on the hydration level. 
The approach to the glass transition of confined water 
is investigated on lowering hydration and 
on supercooling in the framework of Mode Coupling Theories.
At higher hydrations two quite distinct subsets
of water molecules are detectable. Those belonging to the first
layer close to the substrate suffer a severe slowing down, 
while the remaining ones display a scenario typical of 
supercooled liquids approaching the kinetic glass transition. 
}}
\end{flushright}
\bigskip
\baselineskip12.5pt
\noindent
{\bf 1. INTRODUCTION}

\bigskip
\noindent
The study of the modification of the properties of confined water with respect
to the bulk is highly interesting
since in many technological and biological applications water
is confined in porous media.
Several
experimental studies and computer simulations
give evidence that
the perturbation of the substrate and
the geometrical confinement change the properties at 
freezing~\cite{morishige,koga1},
the mobility~\cite{rossky1,lynden-bell} and the dynamical 
behaviour of 
water~\cite{zanotti,philmag}. Particularly attracting is the study of
confined water in the supercooled region.
It is well known that below $235 K$ the crystallization process
driven by the homogeneous nucleation prevents the observation
of a transition to a glass phase of water~\cite{angell}.  
Experimentally forbidden regions of the phase 
diagram of bulk water could become accessible through the study of 
confined water.  
Molecular dynamics (MD) can be a suitable tool for exploring
the approach of water to vitrification. In recent years
MD simulation of Simple Point Charge/
Extended SPC/E~\cite{bgs87} water model potential in the supercooled
phase found a kinetic glass transition at a critical temperature $T_C$,
as defined in the Mode Coupling Theory (MCT), which is
$T_C\sim T_S$ \cite{gallo}, where $T_S$
is the singular temperature of water~\cite{SpeedyAngell}, 
which is $T_S=228K$ or, for SPC/E, 
$49$ degrees below the temperature of maximum density. 
In many experimental studies \cite{zanotti} on confined or
interfacial water a slowing down of
the dynamics of the liquid with respect to the bulk water 
has been found with inelastic
neutron scattering and NMR spectroscopy .  
It has been inferred from experiments that confining water could be 
equivalent to the supercooling 
of the bulk \cite{chen,zanotti}. 
Reduced self diffusion coefficient of water in
contact with solid hydrophilic surfaces, when compared to bulk water 
are found in some of the computer simulation studies~\cite{rossky1} 
but it is still difficult to find
general trends in the
dynamical properties of confined water and there are not any 
systematic studies in the supercooled region.
Among the different systems studied experimentally 
water confined in porous Vycor glass is one of the most interesting
with relevance to catalytic processes and enzymatic
activity. Vycor is a porous silica glass with 
a quite well characterized structure of cylindrical pores
and a quite sharp distribution of the pore sizes around the average value of
$\sim 40 \pm 5 \AA $. Several experiments on water-in-Vycor
have been performed \cite{zanotti,mar1}.
We present here  results obtained from MD simulations of
the single particle dynamics  
of SPC/E water confined in a cylindrical silica cavity
modeled to represent the average properties of Vycor pores
\cite{jmliq1}. 
The SPC/E
potential used for water molecules models a single 
water molecule as a rigid set
of interaction sites with an OH distance of $0.1$~nm and
a $HOH$ angle equal to the tetrahedral angle $109.47^o$.
The Coulomb charges are placed on the atoms. 
Oxygens atoms additionally interact via a Lennard-Jones potential.
SPC/E is particularly suitable for
the study of dynamics in the supercooled region since
it has been explicitly parametrized to reproduce not only the density 
but also the experimental value of the self diffusion constant at ambient
conditions and moreover it is able to reproduce the temperature of 
maximum density of water. 
In our study we will focus on the 
dynamics of confined water at room temperature
where both on lowering the hydration and on supercooling 
we observe a splitting of the time scale in the
relaxation laws.

\bigskip
\noindent
{\bf 2. SLOW DYNAMICS OF WATER CONFINED IN VYCOR}

\bigskip
\noindent
In our simulation we build up a cubic cell of silica
glass of $71 \AA$ with a cylindrical cavity of $40 \AA$ of diameter
as described in detail in previous works~\cite{jmliq1}.
The inner surface of the cylinder is then corrugated by removing
all the silicon atoms bonded to less than four oxygens.
The oxygens dangling bonds are then saturated with acidic hydrogens
in analogy with the experimental preparation of the sample of
Vycor before hydration. 
Water molecules interact with the substrate atoms using an
empirical potential model~\cite{jmliq1,brodka}. The periodic 
boundary conditions are applied along the 
axis of the cylinder. 
\begin{figure}[t]
\unitlength1cm
\begin{minipage}[t]{8.7cm}
\includegraphics[width=8.7cm,height=9cm]{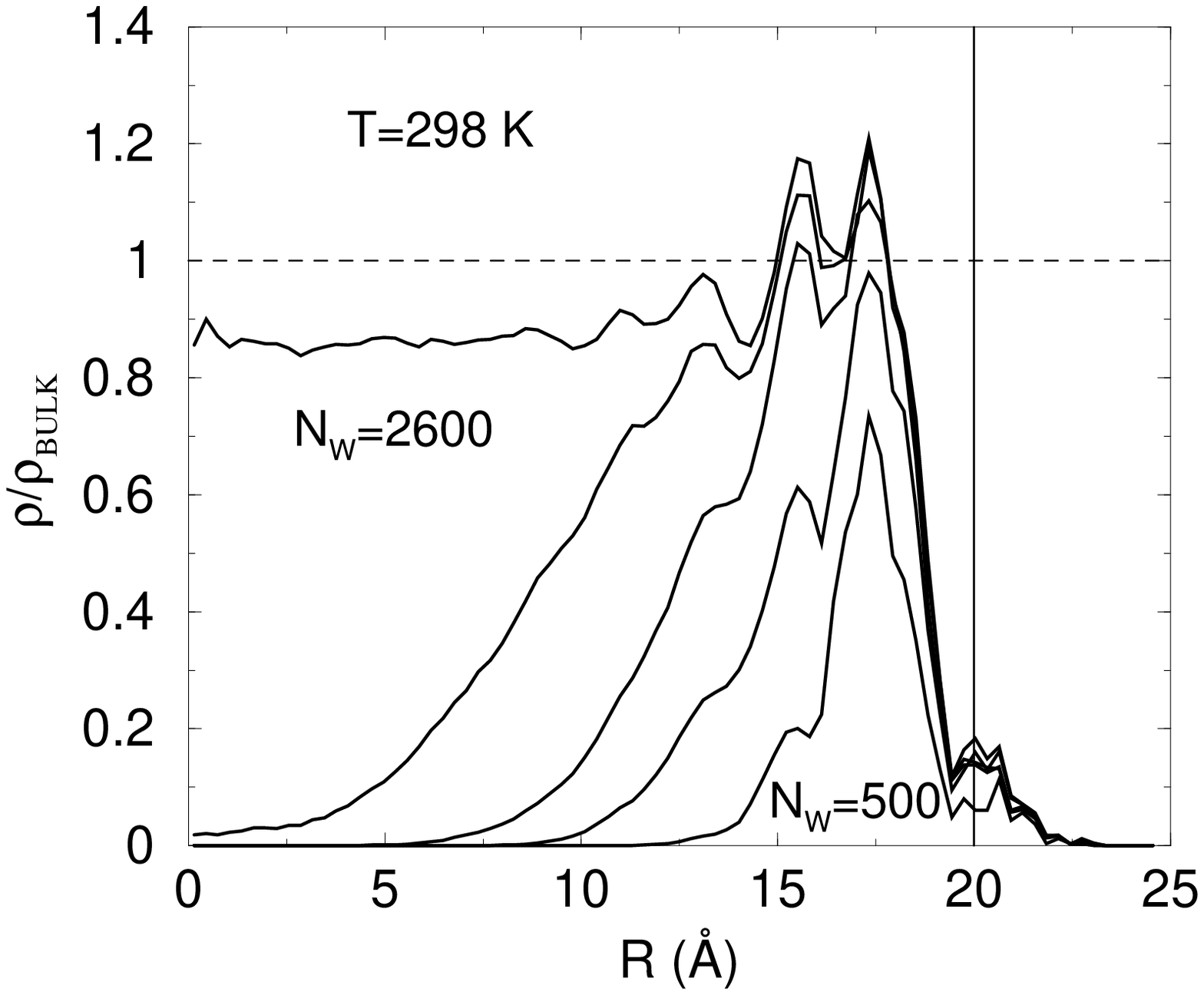}
\parbox{8cm}
{\footnotesize {\bf Figure~1.~}
Radial density profiles 
normalized to the bulk 
for oxygen atoms at ambient temperature 
for the hydration levels investigated (see text). 
Curves on the top correspond to higher levels of
hydration.}
\end{minipage}
\begin{minipage}[t]{8.7cm}
\includegraphics[width=8.7cm,height=9cm]{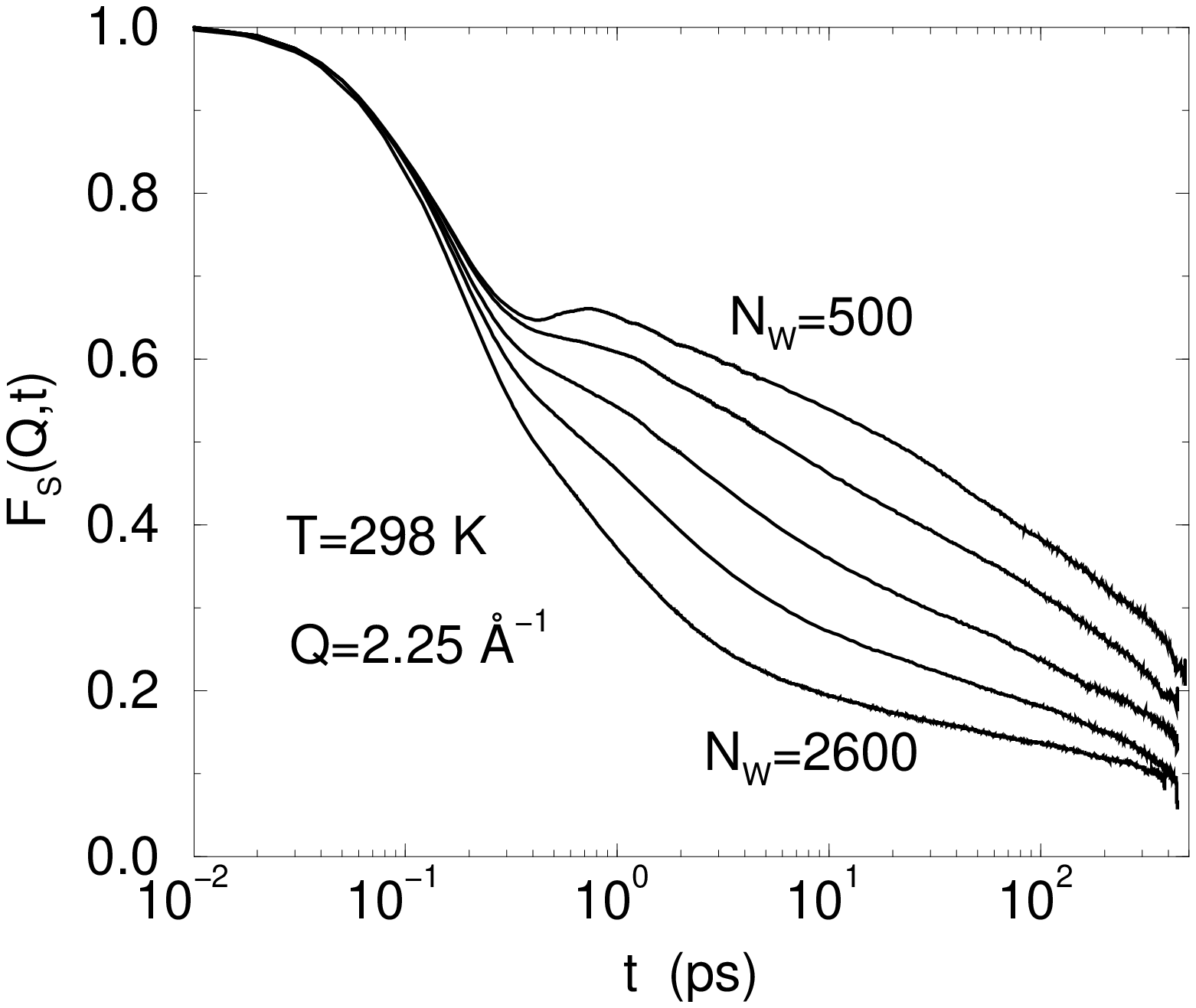}  
\parbox{8.2cm}
{\footnotesize {\bf Figure~2.~}
Self intermediate scattering function of oxygens for
ambient temperature at the peak of the structure factor
for the hydration levels investigated.
Curves on the top correspond to lower levels of hydration.}
\end{minipage}
\end{figure}
\noindent

The molecular dynamics calculations have been performed for different
numbers of SPC/E water molecules, 
introduced in the pore, corresponding to different levels of
hydration:
$N_W=2600$ corresponding to $96\%$ of hydration level of the pore,
$N_W=2000$ ($74\%$),
$N_W=1500$ ($56\%$), $N_W=1000$ ($37\%$) and $N_W=500$ ($19\%$).
We will discuss here the results obtained for two temperatures:
$T=298$~$K$ for all the hydrations and $T=240$~$K$
for $N_W=1500$. A more detailed report will be published 
elsewhere~\cite{future}.
As a first result we show in Fig.1 the density profiles
along the pore radius. 
Already at lower hydrations the
presence of a layer of water molecules wetting the substrate surface
is observed.
At nearly full hydration two layers of water with higher than bulk
density are evident.
There is a strong tendency of water molecules
close to the surface to form hydrogen bonds
(HB) with the atoms of the substrate~\cite{jmliq1}.

In the MCT~\cite{goetze} description of a liquid approaching the 
glass transition 
the dynamic behaviour is mastered by the  ``cage effect''.   
The molecule is 
trapped by the transient cage formed by its nearest neighbours.
Signatures of the MCT behaviour  can be 
found in the intermediate scattering function, ISF
$F_S(Q,t)$, which will be shown below for our system.
We expect to observe a diversification of the relaxation times
resulting in a development of a shoulder in the relaxation laws.
The long time tail of the ISF is predicted
to have a stretched exponential behaviour when the system approaches the
glass transition in the framework of MCT.
In Fig.2  we present the ISF for the water oxygens
at the 
oxygen-oxygen peak of the structure factor. 
The ISF are displayed at room temperature
for the different hydrations.  
Upon decreasing hydration level the development of the
shoulder in the relaxation laws is evident.
\begin{figure}[h]
\unitlength1cm
\begin{minipage}[t]{8.7cm}
\includegraphics[width=8.7cm,height=9.0cm]{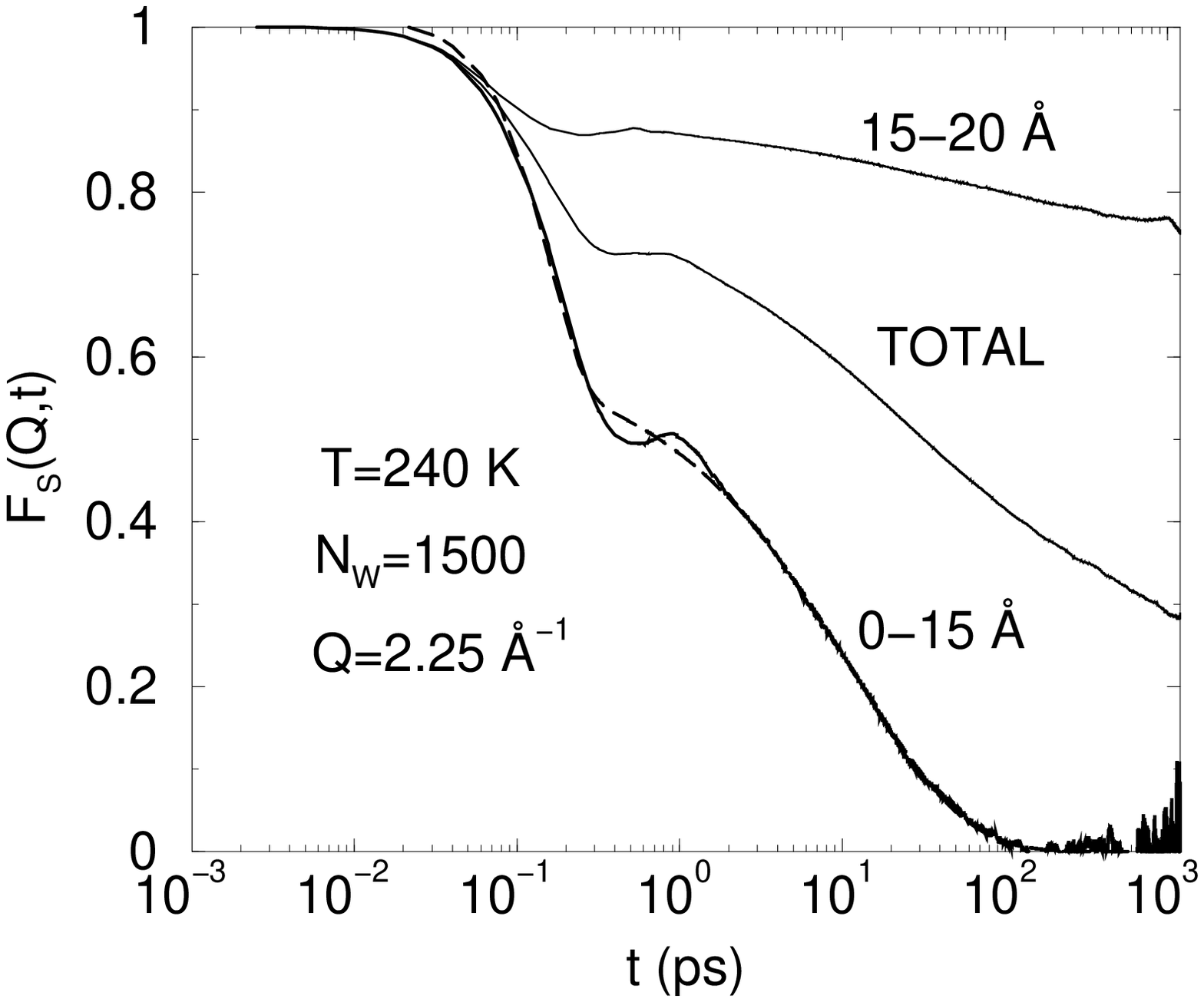}
\parbox{8cm}
{\footnotesize {\bf Figure~3.~}
Shell analysis of the self intermediate scattering function (ISF)
for oxygens at~$T=240~K$ and~$N_W=1500$ (roughly half hydration).
The curve on the top corresponds to the contribution to the ISF
coming from the water molecules moving in the two layers closer
to the substrate (outer shells), see Fig.1. 
The curve on the bottom (continuous line) 
represents the contribution of the remaining 
water molecules that move closer to the center of the pore (inner shells).
The dashed line is the fit to eq.\protect\ref{strexp}. The central curve
displays the total contribution.}
\end{minipage}
\begin{minipage}[t]{8.7cm}
\includegraphics[width=8.7cm,height=9.0cm]{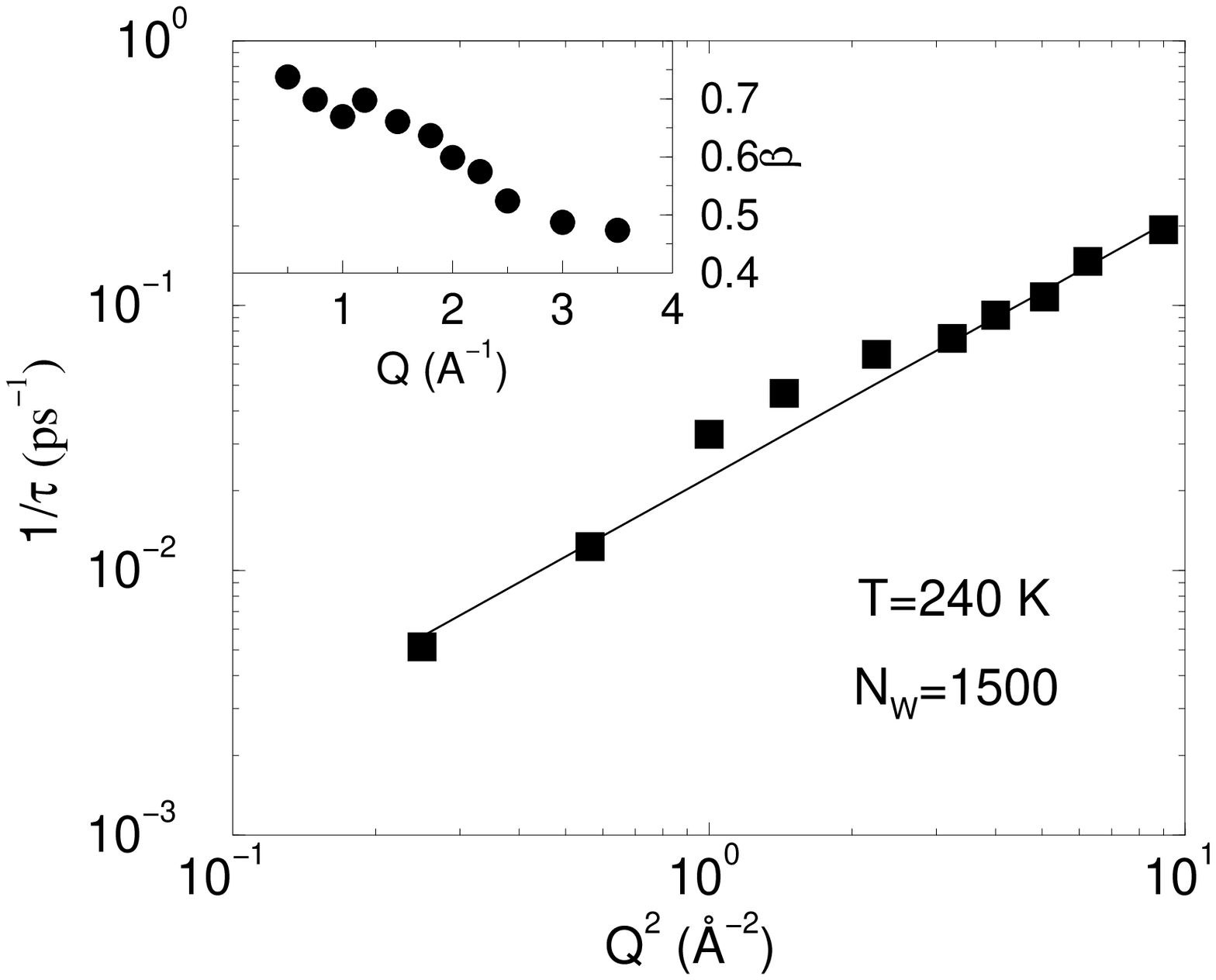} 
\parbox{8.2cm}
{\footnotesize {\bf Figure~4.~}
Values of $\tau_l$ vs. $Q^2$ (main picture)
and $\beta$ vs $Q$ (inset) extracted ~from~the~fits~ of the 
intermediate scattering function to eq.\protect\ref{strexp} (see~text~).
The line with slope 1 (continuous line) in the main picture
has been drown as a guideline for the eyes.}
\end{minipage}
\end{figure}

All the tails of the correlators of Fig.2
are highly non-exponential. None-the-less
these ISF could not be fitted
to the same formula used for bulk supercooled water
\cite{gallo}, 
\begin{equation}
F_S(Q,t)=
\left[ 1-A(Q) \right] e^{-\left( t/\tau_s \right)^2}+
A(Q)e^{- \left( t/\tau_l \right)^\beta}
\label{strexp}
\end{equation}
where $A(Q)=e^{-a^2Q^2/3}$ is the Lamb-M\"ossbauer factor 
(the analogous of the Debye Waller Factor for the single particle)
arising from the cage effect, 
$\tau_s$ and $\tau_l$ are, respectively, the short and
the long relaxation times and $\beta$ is the Kohlrausch exponent.
Due to the strong hydrophilicity of the pore we observe a diversification of 
dynamic behaviour as we proceed from the pore surface to
the  center of the pore.
The function $F_S(Q,t)$  is therefore splitted
into the contribution coming from the two layers of water 
molecules closer to the pore surface (outer shells)
and  
the contribution coming from all the remaining ones (inner shells)
as displayed in Fig.1.
The shell contribution is given only by the particles that
move in the selected shell.
We find that the inner shell contribution could be 
perfectly fit to eq.\ref{strexp},
while the outer shells one decays to zero over a 
much longer timescale so that water molecules there behave already as a 
glass. 

The double step relaxation observed for the lowest hydration deserves
some comment. In fact for $N_W=500$ the surface coverage is not
complete and patches of water molecules are visible along the pore
surface. The MCT is able to account also for the dynamics of
clusters in a frozen environment~\cite{goetze} and this might be
the reason of the development of the diversification of the relaxation
law. From our analysis of ISF turns out that the behaviour of molecules
belonging to the first hydration layers changes as a complete surface
coverage is achieved.

In Fig.3 we show an example of the shell analysis 
for T=240 K and $N_W=1500$. The
top curve is the contribution to the total ISF coming from the molecules 
in the two shells closest to the substrate. 
The bottom curve is the contribution
from the molecules in the remaining shells. The central one is the total ISF. 
The dashed line is the fit to eq.\ref{strexp}. From the fit we obtain
$\beta=0.62$, $\tau_l=11$ $ ps$ $\tau_s=0.16$ $ps$. 
The  $\tau_l$ is similar to that of bulk water at the same
temperature while the $\beta$ is much lower. 
For the inner shells a bump around $0.7 ps$  is also observed that 
could possibly be related to the existence of the  Boson Peak feature 
in the $S(q,\omega)$ \cite{binder}. This bump also appears in Fig.2
for the lowest hydration level investigated, namely $19$\%.
In Fig.4 the $\tau_l$ and the $\beta$ values extracted from the
fits to eq.\ref{strexp} are plotted as a 
function of $Q$ for T=240 K and $N_W=1500$.
The $\beta$ value reaches a plateau value and the $\tau_l$ values show a
$Q^2$ dependence. Both these behaviours are found in some glass former 
undergoing a kinetic glass transition and in particular the $Q^2$
behaviour has been observed
for example in glycerol close to the
glass transition \cite{glice}. However other behaviours of the relaxation
times as a function of $Q$ have been also found \cite{colme}.

\bigskip
\noindent
{\bf 3. CONCLUDING REMARKS}

\bigskip
\noindent
We report some of the results achieved in the computer 
simulation of the dynamics of water molecules confined
in a silica pore. The substrate is carefully modeled 
to reproduce the main features of the hydrophilic cavity of
Vycor glass. 
In the water density profile we observe at high hydration levels
a double layer structure which strongly influences the dynamics
of the molecules.

An analogy between supercooled bulk water 
and confined water as a function of hydration 
level of the pore is possible in the sense that
upon decreasing the hydration level a glassy behaviour 
appears already at ambient temperature.
None-the-less the approach of
confined liquid to the kinetic glass transition is rather
different with respect to its bulk phase. 
In particular  
two quite distinct subsets of water molecules are detectable
for confined water. The subset 
that is in contact with the surface is at higher density
with respect to the bulk and is already a glass with low mobility
even at ambient temperature. The inner subset displays, like
a supercooled liquid, a two step relaxation behaviour.
The shape of the ISF long time tail, also called late part
of the $\alpha$-relaxation region, can be perfectly fitted to
a stretched exponential function. The $\beta$ and $\tau_l$ behaviours
are consistent with the values extracted for other glass formers 
in literature.

More simulations on this system as function of temperature
and hydration are in progress for a full
MCT test and a complete comparison with the bulk. 
The relation between the two types of subsets and the so called
{\it free} and {\it bound} water found in several experiments
on confined water at freezing~\cite{morishige} is worthwhile to be explored
in the future.

\bigskip

\noindent
{\bf Acknowledgments}
\bigskip

\noindent
The authors wish to thank M.A. Ricci and E. Spohr for their contribution
to this work. 


\end{document}